**The Effects of Rotation Rate on Deep Convection in Giant Planets with Small Solid Cores**

Martha Evonuk* and Gary A. Glatzmaier

*Department of Earth Sciences, University of California, 1156 High St., Santa Cruz, CA 95064 USA*

*Corresponding author. E-mail: mevonuk@es.ucsc.edu, Fax: 831-459-3074

Now at: Institut fuer Geophysik, ETH Zuerich, Switzerland

**Abstract**

We study how the pattern of thermal convection and differential rotation in the interior of a giant gaseous planet is affected by the presence of a small solid core as a function of the planetary rotation rate. We show, using 2D anelastic, hydrodynamic simulations, that the presence of a small solid core results in significantly different flow structure relative to that of a fully convective interior only if there is little or no planetary rotation.

**1. Introduction**

Rotation significantly affects the internal structure and dynamics of giant gaseous bodies such as giant planets and stars. The vorticity generated by fluid rising or sinking through the density-stratified interior of a planet or a star and the resulting nonlinear convergence of angular momentum flux maintains differential rotation in the interior and zonal winds at the surface (Evonuk and Glatzmaier, 2006; Glatzmaier et al., in preparation). The strength and pattern of the resulting shear flow depends on the rotation rate and the rate of internal heating, that is, the strength and nature of convection.

A related feature of interest concerns the presence of small solid or otherwise non-interacting cores in these bodies. In particular, does the fluid behavior change significantly with the presence of a small non-convecting core? This question is of interest for two reasons. First, it is likely that a significant percentage of extra-solar gaseous planets have fully convective interiors. Giant planets are believed to have formed either through accretion of a solid core followed by the capture of surrounding gaseous hydrogen and helium (Pollack et al., 1996) or possibly by direct gravitational instability of the gas in the disk (Boss, 1998). After the giant planets form their cores may erode. Core erosion can occur if the heavy elements are at least partially soluble in the hydrogen helium envelope and if there is sufficient convective energy to overcome the molecular weight barrier that is created (Stevenson, 1982). However, initially high central temperatures (~30,000K) favor solubility of the heavy elements (Guillot, 2005) and could allow significant core erosion via convection or tidal interactions and subsequently result in fully convective planets (Guillot et al., 1994). In our own solar system, one-dimensional structure models of Jupiter indicate that its core must be less than 10 Earth masses; it has no core in extreme models that assume a particular hydrogen EOS, a large $J_4$ value and an enhancement in heavy elements by 4 to 6 times the solar value (Guillot, 1999). A second reason to study the behavioral differences between fluid bodies with and without non-convecting cores is that most three-dimensional, global, hydrodynamic models of stars and giant gaseous planets, because of the numerical methods, are not able to simulate convection without a solid core (e.g., Glatzmaier, 1984; Sun et al., 1993; Aurnou and Olson, 2001; Christensen, 2001; Brun et al., 2003; Kuhlen et al., 2005). Consequently it is important to know if the presence of a non-convecting core causes a significant change in the fluid flow patterns, relative to a fully convective planet, at various rotation rates.

We compare via 2D numerical simulations fluid behavior in giant planets with rapid rotation (period of 10 hours), with slow rotation (2 months), and in the extreme case with no rotation for identical size, mass, and heating rates. We show that at high rotation rates the case with a small core is essentially equivalent to that with no core; and conclude that models that are unable to remove the solid core produce solutions that are reasonable approximations for cases without a core. However at low rotation rates, the geometry of the simulation becomes more important. The fluid in a body with no core develops a dipolar flow structure. This differs significantly from the multi-cell structure formed in a body with a small core. Therefore, models that include a solid, non-convecting core are inappropriate for studying slowly rotating fully convective bodies, such as tidally locked planets in orbits close to their host star or fully convective A-type stars.

## 2. The Numerical Method

The test cases are two-dimensional (2D) time-dependent nonlinear computer simulations of thermal convection in the equatorial plane of a planet with central gravity. We chose to model the planets in two dimensions to provide higher resolution and obtain

69 more turbulent flow. Of course this advantage is offset by the fact that 2D simulations
70 can not capture all the internal dynamics of a 3D fluid sphere.

71 The following anelastic equations of momentum (1), heat (2), and mass
72 conservation (3) are solved with the finite volume method on a Cartesian grid as in
73 Evonuk and Glatzmaier (2006):

74 $$\frac{\partial}{\partial t}(\bar{\rho}u) = -\nabla \cdot \left[\bar{\rho}u_i u_j - P\delta_{ij} - 2\nu\bar{\rho}\left(e_{ij} - \frac{1}{3}\nabla \cdot u \delta_{ij}\right)\right] - \rho\bar{g}\hat{r} + 2\bar{\rho}u \times \Omega - \rho\Omega \times (\Omega \times r), \quad (1)$$

75 $$\frac{\partial}{\partial t}(\bar{\rho}S) = -\nabla \cdot \left[\bar{\rho}(S+\bar{S})u - \kappa_t \bar{\rho}\nabla S - \frac{C_p \kappa_r \bar{\rho}}{\bar{T}}\nabla T\right] + \frac{\bar{\rho}}{\bar{T}}\frac{d\bar{T}}{dr}\left[\kappa_t \frac{\partial S}{\partial r} + \frac{C_p \kappa_r}{\bar{T}}\frac{\partial T}{\partial r}\right] + \bar{\rho}Q_s, \quad (2)$$

76 and $\nabla \cdot (\bar{\rho}u) = 0$, (3)

77 where $\rho$ is the density, $u$ is the velocity vector, $P$ is the pressure, $\nu$ is the turbulent viscous
78 diffusivity, $e_{ij} = \frac{1}{2}\left[\frac{\partial u_i}{\partial x_j} + \frac{\partial u_j}{\partial x_i}\right]$ is the rate of strain tensor, $g$ is the gravity, $\Omega$ is the rotation
79 vector, $S$ is the entropy, $\kappa_t$ is the turbulent diffusivity, $C_p$ is the specific heat capacity at
80 constant pressure, $T$ is the temperature, $\kappa_r$ is the radiative diffusivity, and $Q_s$ is the heating
81 function. The over-barred quantities are prescribed reference state variables that are
82 functions of only radius, and the viscous and thermal diffusivities are constant.

83 Our model is a simple density-stratified fluid spanning four density scale heights,
84 which places our outer boundary at 8.9 kilobars if we were modeling Jupiter (Guillot,
85 1999). It is important to include this density variation not only because density-stratified
86 fluids behave differently than constant-density fluids (Evonuk and Glatzmaier, 2004), but
87 also because density-stratification is needed in 2D for the Coriolis force to influence the
88 fluid flow (Glatzmaier et al., in preparation). We set the turbulent Prandtl number ($\nu/\kappa_t$)
89 in the convective zone to 0.1 and the traditional Rayleigh numbers for these cases to a

90  few times $10^{12}$. The spatial resolution is 1000x1000 in Cartesian coordinates. **While the finite-volume method in this coordinate system eliminates the singularity at the center of the disk and allows for nearest neighbor communication (and therefore efficient parallelization), it is also much more diffusive than the spectral methods usually used and therefore requires more grid points than spectral modes to obtain comparable accuracy. High resolution also improves the "roundness" of the boundaries and allows the resolution of smaller scale features embedded in the larger scale flows.** Each case was run for more than 500,000 numerical time-steps corresponding to 0.002 viscous diffusion times or approximately 200 convective turnover times.

The non-convecting cores extend to 10% of the total radius ($r_t$). The outer 10% of the disks are stable radiative layers providing a buffer between the outer impermeable boundary and the internal convecting fluid. The inner 35% of the disks are heated to drive convection (Evonuk and Glatzmaier, 2006). **The radial heating profile is a fairly arbitrary choice and mainly meant to maintain convection. Stars of course are heated in their central cores and fluid planets can generate heat from gravitational contraction, which is proportional to pressure. However, this latter effect may not be significant in planets with electron degeneracy.** The fast rotator spins at about Jupiter's rotation period of 10 hours, whereas the slow rotator has a two-month period. The Ekman number, the ratio of the viscous forces to the Coriolis force, $Ek = \frac{\nu}{2\Omega D^2}$, is 3.0x$10^{-8}$ for the high rotation cases and 4.4x$10^{-6}$ for the low rotation cases.

**3. Results and Discussion**

At higher rotation rates it is difficult to distinguish between the cases with small and no core. Figure 1a and 1b show snapshots of the entropy perturbation overlaid with velocity arrows as viewed from the north with the rotation vector out of the page. The fluid flow is clearly differentiated into prograde motion in the outer part of the disk and retrograde motion in the inner part for both cases, as can also be seen in the plot of the zonal flow with radius (Figure 1c). This differential rotation is maintained by the generation of vorticity as fluid flows through the density stratification. The Coriolis force generates negative vorticity as hot material expands as it rises from the interior of the disk. Likewise sinking material generates positive vorticity in the outer regions of the disk. **One needs to distinguish between the local nonaxisymmetric vorticity generation by rising and sinking plumes and the global axisymmetric zonal flow, which is maintained by the convergence of the nonlinear product of nonaxisymmetric radial flow and nonaxisymmetric longitudinal flow. Part of this nonlinear product has an axisymmetric eastward (i.e., prograde) component, which accumulates in the outer part of the convection zone, and an axisymmetric westward (retrograde) component, which accumulates in the inner part of the convection zone. This occurs because plumes sinking from near the surface curve westward (i.e., transport westward momentum away from the outer region). We focus on sinking plumes near the surface because this mechanism is most effective where the density scale height is smallest and the radial velocity is the greatest.** The resulting fluid trajectories cause a nonlinear convergence of prograde momentum in the outer part and retrograde momentum in the inner part (Glatzmaier et al., in preparation). The suppression of large-scale convection shifts the peak kinetic energy to higher wave numbers, as oscillations, instead of traditional convection cells, develop to transport the thermal energy vertically (Figure 1d). These small-scale features, as well as the effect of a larger non-convecting core, are discussed in more detail in Evonuk and Glatzmaier (2006).

Removing rotation from the simulations eliminates the effect of the Coriolis force while retaining the effect of the density stratification on convection. Now we see distinctly different flow patterns for the cases with and without solid cores (Figure 2a,b). The case without a solid core forms two nearly symmetric cells with peak flow velocities in the center of the disk which is the most efficient flow pattern for removing the heat generated in the central region. The symmetry of these two cells results in mean zonal flows close to zero through out the disk (Figure 2c). The peak kinetic energy is at wave number one reflecting the dipolar flow structure (Figure 2d). The pattern of convection we see is similar to the lowest symmetric mode ($l = 1$, $m = 0$) of disturbance analytically predicted by Chandrasekhar (1961) in a fluid sphere. Conversely, in the case with a small core, the core breaks up the flow through the center into smaller cells (Figure 2b). This produces higher kinetic energies throughout the energy spectrum for all but wave number one (Figure 2d). The mean zonal flow in this non-rotating case should average to zero over long times. The prograde zonal flow near the core in Figure 2c is due to the persistence of a dominant cell (Figure 2b) during the time over which this average was made. We expect that in a longer simulation that large cell would eventually dissipate and another would form with the opposite vorticity.

In the intermediate case with slow rotation we see convective cell patterns similar to the case with no rotation (Figure 3a,b). However, the non-zero Coriolis forces now organize the flow to maintain a weak differential rotation, prograde in the outer part and retrograde in the inner part, as in the high rotation case (Figure 3c). Although the two cases behave differently at depth they have similar zonal velocities at the base of the radiative zone. This suggests that surface observations of the zonal wind velocities would be indistinguishable. However, magnetic fields generated in the deep interior may have different structures at the surfaces of slowly rotating giant planets with and without small solid cores.

**4. Summary**

This series of simulations illustrate the increasing dominance of the Coriolis effect as the rotation rate increases at a given heating rate, suppressing convection and establishing differential flow in the planet. At high rotation rates, $Ek<10^{-8}$ for our 2D simulations, the presence of a small non-convecting core has little effect on the structure of the flow. However, in more slowly rotating planets ($Ek>10^{-5}$ in 2D), the presence of a non-convecting core impedes the tendency for dipolar flow through the center, a potentially important effect in giant planets and stars with slow rotation rates especially in light of possible magnetic field generation and interactions in these bodies. **Similar to the 2D results, preliminary low resolution 3D finite-volume anelastic simulations without a core show dipolar flow in cases with low rotation rate while cases with higher rotation rates have essentially no flow through the center as the fluid circulates about the origin instead of flowing through it.**

**Acknowledgements**

Support provided by NASA Planetary Atmospheres Program (NAG5-11220), NASA Outer Planets Program (OPR04-0017-0109), and the Institute of Geophysics and Planetary Physics at the Los Alamos National laboratory and the University of California, Santa Cruz. Computing resources were provided by an NSF MRI funded Beowulf cluster at UCSC and allocations at the Pittsburgh Supercomputing Center, National Center for Atmospheric Research, San Diego Super Computing Center, and NASA Ames.

**Figure Captions**

Figure 1: Snapshots at high rotation rate (ten-hour period) of the entropy perturbation overlaid with velocity arrows for the case without a solid core (a) and the case with a solid core (b). High entropy is indicated with lighter shades of red and low entropy with

228  darker shades and the rotation axis is directed out of the page.  The velocity arrows are
229  scaled the same for the two cases and for those in the next two figures.  Also shown are
230  time averaged plots of the mean zonal wind speed as a function of radius (c) and the
231  kinetic energy spectra as a function of longitudinal wave number (d).  The case with no
232  core is shown in these plots with a dashed line.

234  Figure 2:  Same as Figure 1 but for cases with no rotation.

236  Figure 3:  Same as Figure 1 but for cases with slow rotation rate (two-month period).

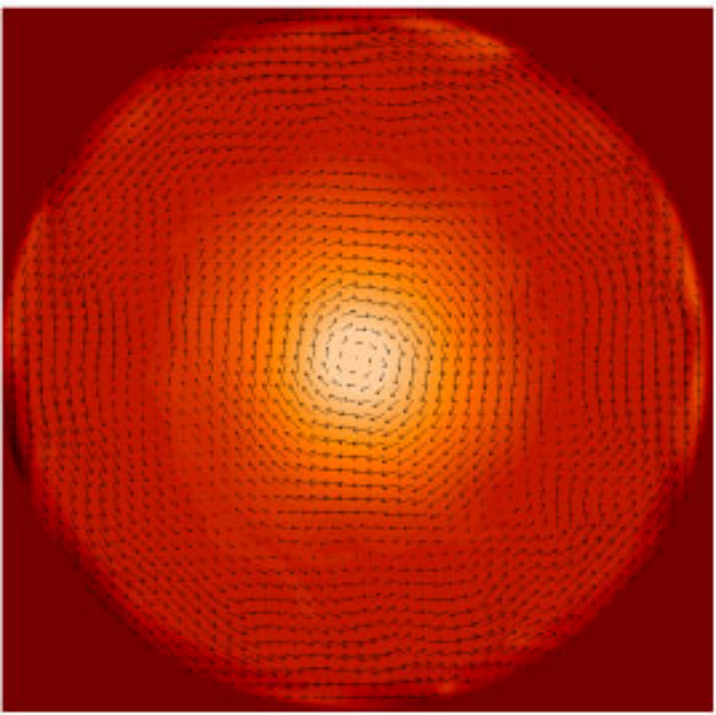
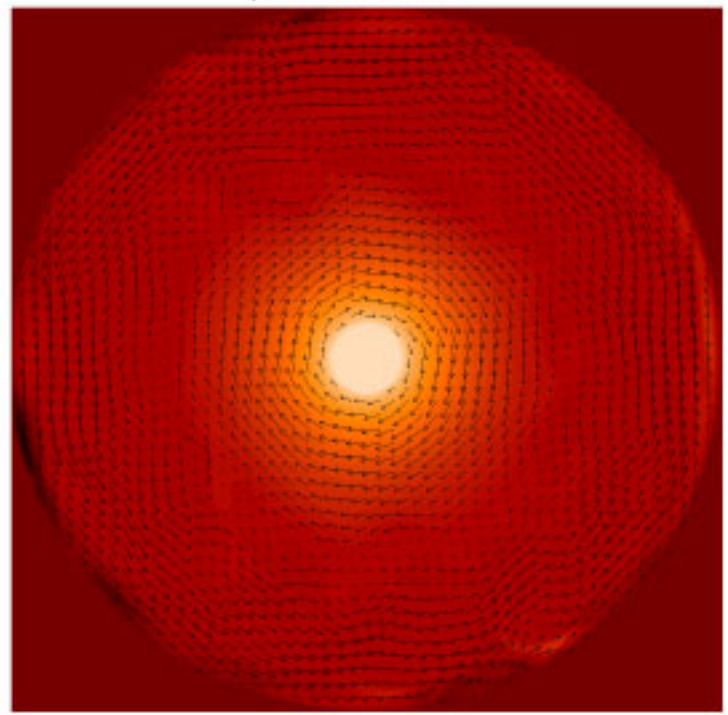
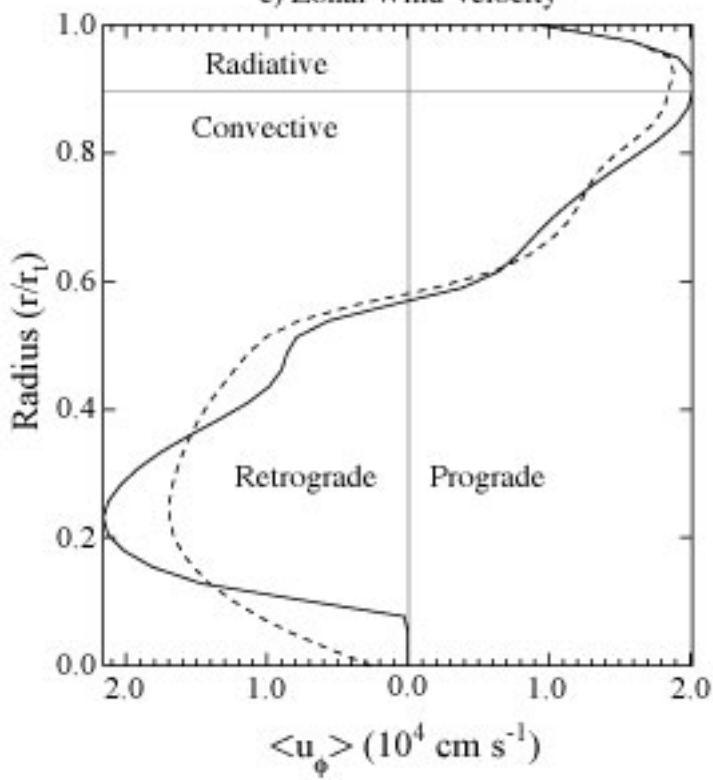
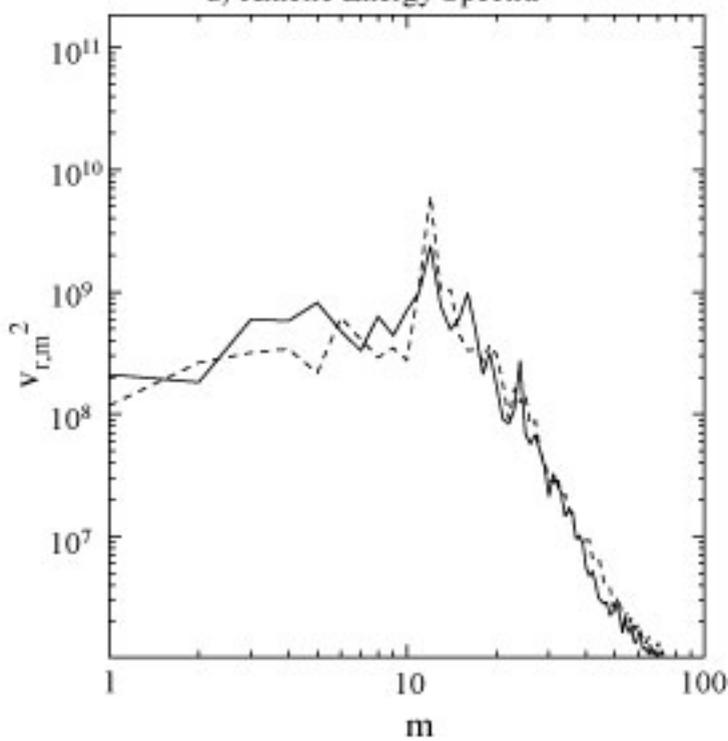

Figure 1: High Rotation Rate

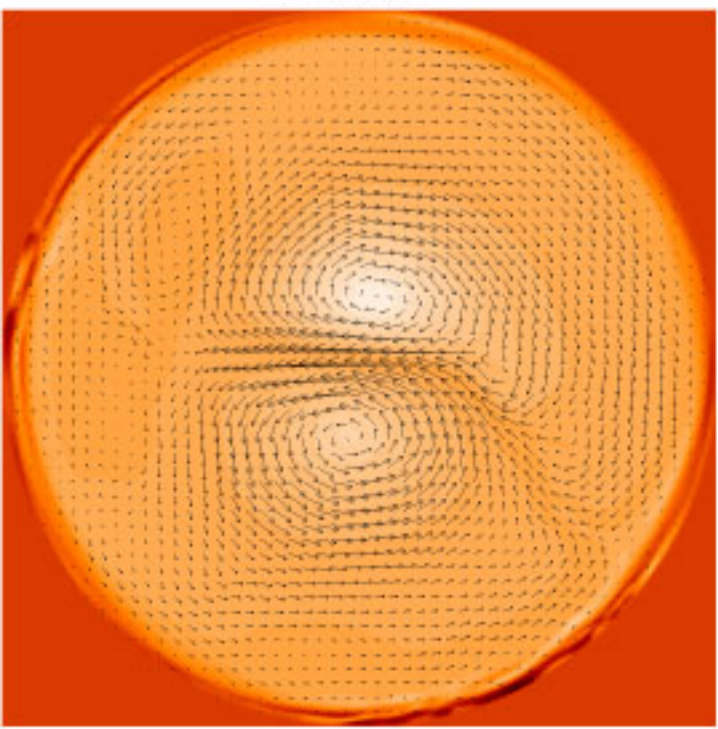
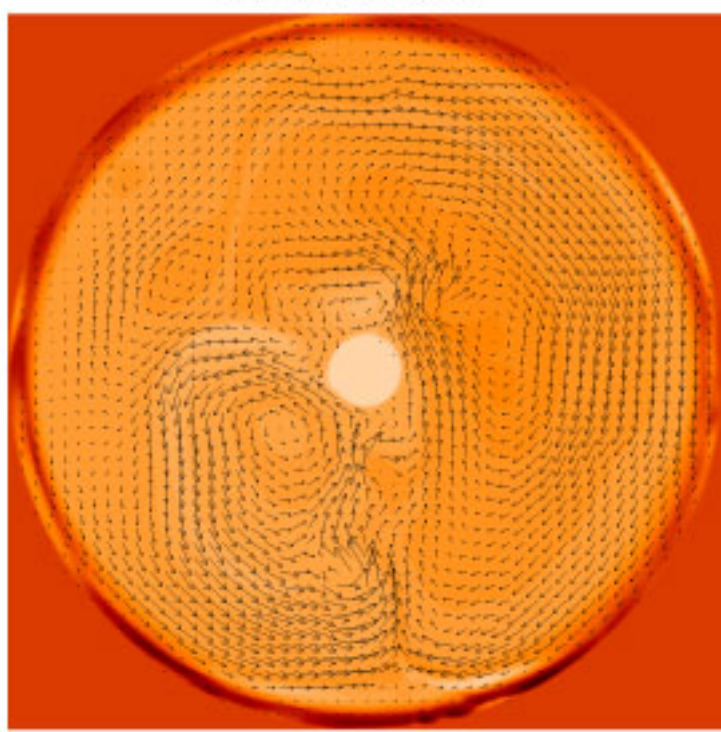
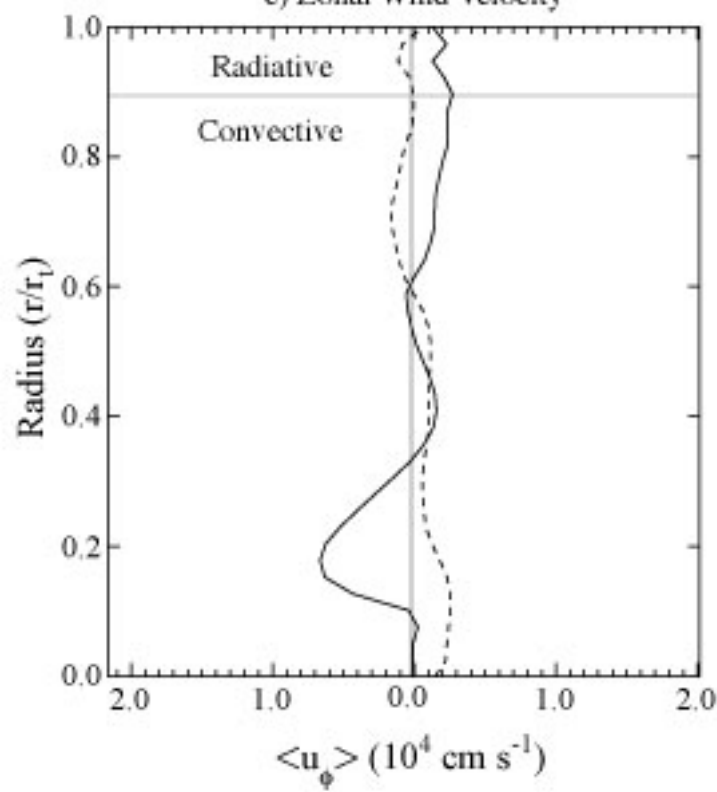
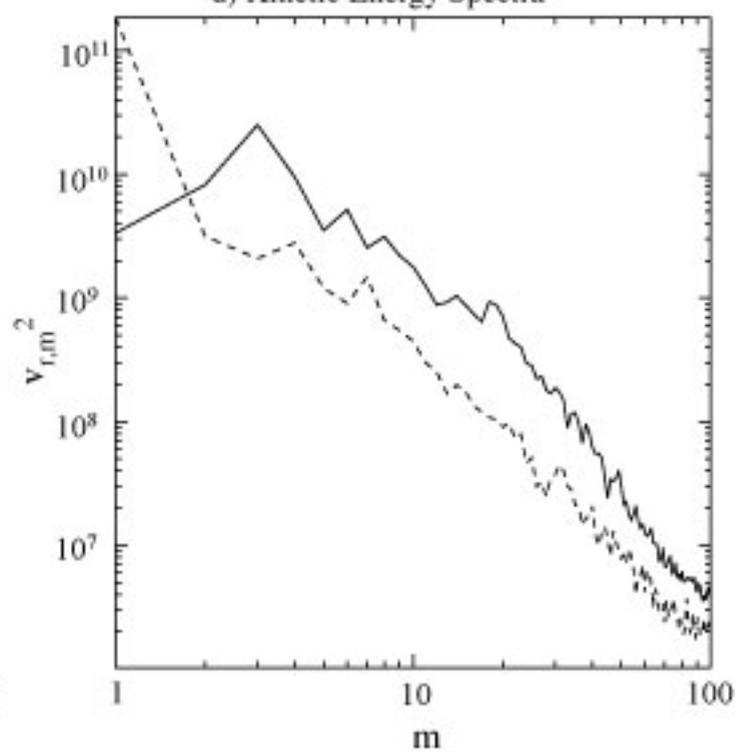

Figure 2: No Rotation

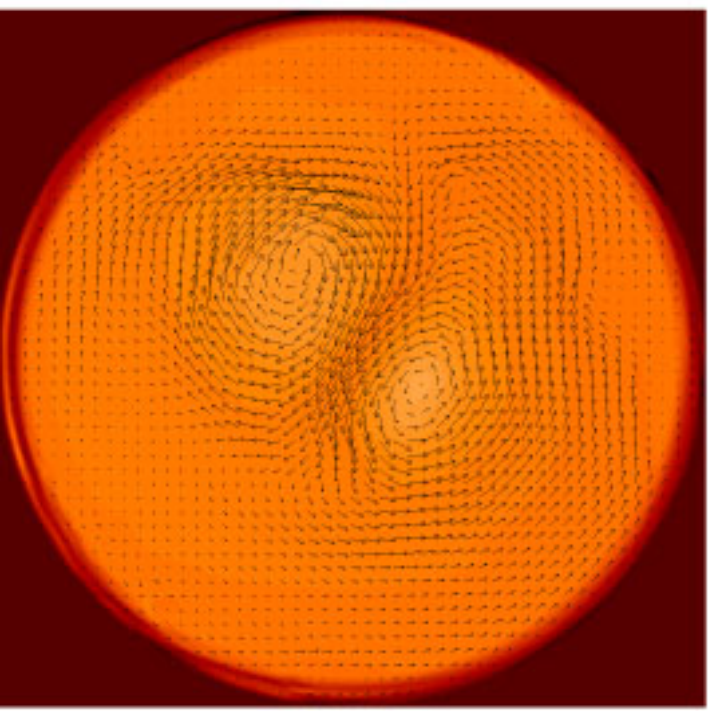 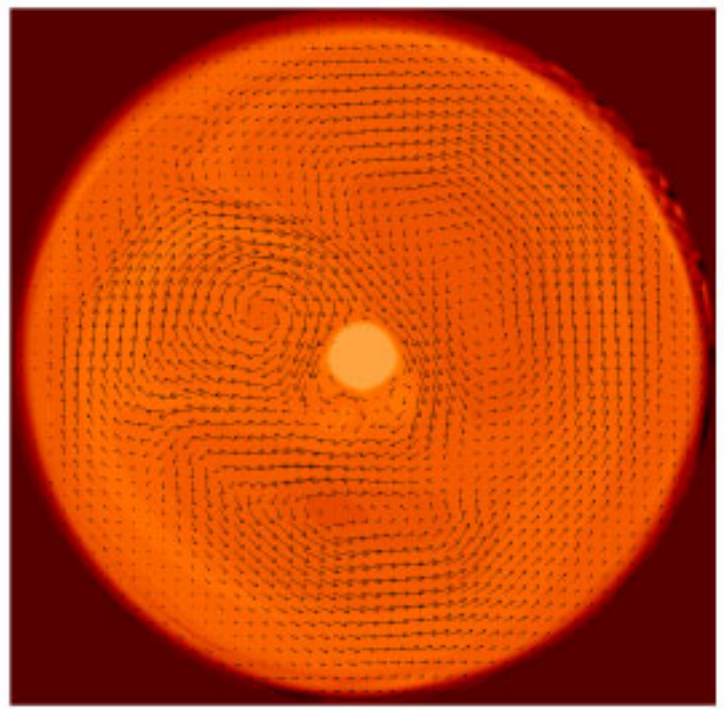
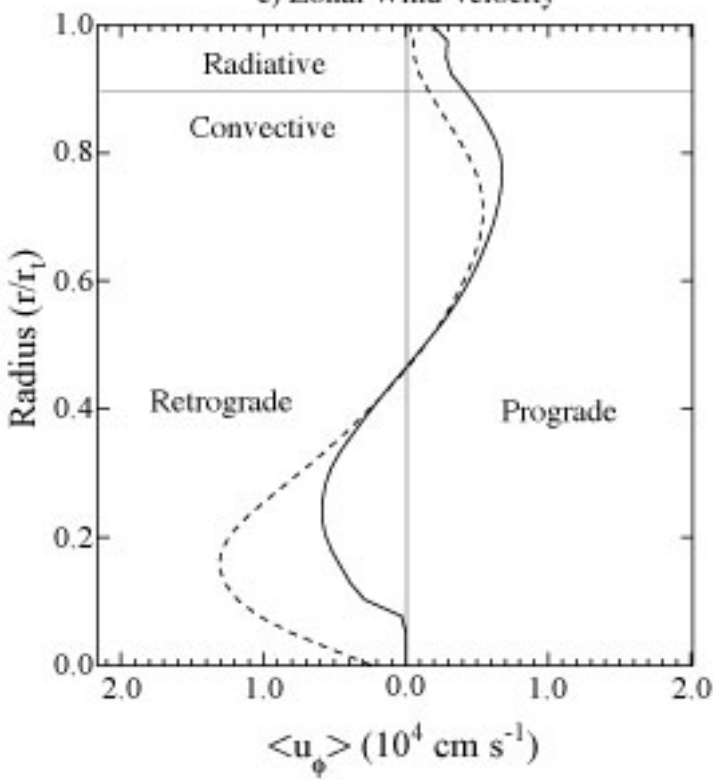 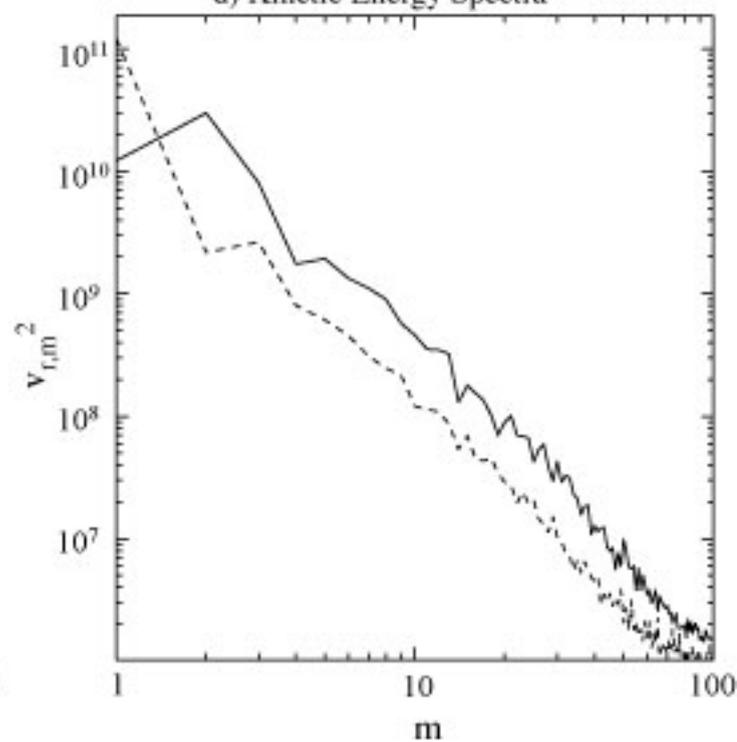

Figure 3: Low Rotation Rate